\documentstyle[12pt]{article}
\begin{document}
\title{Constructive Complexity and Artificial Reality: An Introduction}
\author{
        Kunihiko KANEKO\\
        {\small \sl Department of Pure and Applied Sciences}\\
        {\small \sl University of Tokyo, Komaba, Meguro-ku, Tokyo 153, JAPAN}\\
and \\
          Ichiro TSUDA\\
        {\small \sl Department of Mathematics}\\
        {\small \sl Hokkaido University, Sapporo 060, JAPAN}
\\}
\date{}

\maketitle
\begin{abstract}

Basic problems of complex systems are outlined with an
emphasis on irreducibility and dynamic many-to-many correspondences.
We discuss the importance of a
constructive approach to artificial reality and
the significance of an internal observer.

\end{abstract}

\section{Introduction}

For over hundred years, scientists, above all, physicists have tried to
understand the complexity of nature, by decomposing it into simple processes
that can be dealt with by simple theories.  This strategy is often referred to
as
``Ockham's razor".  It was regarded as
ideal, in modern science, to describe a system in terms of a small number of
parameters, variables, equations, etc.  In a strict sense, however, the
reduction to a system with a small number of degrees of freedom is
not always possible.  Then,
one may introduce a ``noise" term instead, in order to take
residual degrees of freedom into account.

In statistical physics, this reduction was successful because of
the introduction of appropriate order parameters.
Even when a system has many degrees of
freedom, it can often well be described by a macroscopic order parameter
with a corresponding noise term, as can, for example be seen
in equilibrium
statistical mechanics, linear-response theory, system-size (omega)
expansion \cite{van-Kampen,KMK}, and slaving principle \cite{Haken}.

A related paradigm is the use of a ``mode".  Here a system is assumed to be
represented by the superposition of some modes, like the
Fouirer modes.  In solid-state physics
these are attributed to some excitation, termed as ``---on's".
The use of modes is powerful as long as the system can be
approximated by a linear one.
In dynamical systems, it is also successful, even
if the system is nonlinear, as long as it is not chaotic.  It can be
employed, for example, in quasiperiodic motion on a torus and in
the representation by solitons.

The {\sl reductionists'} pictures have been challenged by the discovery
of chaos.  First, the amplification of a tiny perturbation in chaos implies
that
the separation
between microscopic and macroscopic levels is no longer possible.
Second, the picture of ``modes" is not straightforwardly applicable to chaos:
Even if a system has just three degrees of freedom,
it can implicitly include continuously many modes.
For example, a  chaotic system cannot be represented by a finite number of
Fourier modes.
A one-to-one correspondence of simply chosen coordinates
(such as Fourier modes) to original
variables is no longer valid here.  Complexity in
grammatical rules to characterize chaos is discussed in detail by
{\bf  Crutchfield} in the present volume.

Another challenge to the traditional picture can be found
in a system called spin glass,
which originated in
the statistical mechanics of spin system with random interaction \cite{SG}.
In relation with the phase transition problem,  physicists searched for
order parameter(s).  However, a detailed theoretical
analysis shows that the order in the low-temperature phase is represented
only by a functional of order parameters, rather than
a finite number of them.
Indeed, in the neural network model based on the statistical
mechanics of nonhomogeneous spin systems ( often called as Hopfield model), a
one-to-one correspondence
between an interaction code and an attractor is no longer valid,
when the number of stored input patterns is larger than a certain threshold.
In spin glasses, the correspondence between an interaction and an attractor
( or a thermodynamically metastable state) is highly complex,
while, the sample dependence (i.e., dependence on the choice
of couplings) remains finite even in the thermodynamic limit.

Thus the one-to-one correspondence between states and representations
is challenged statically by spin glasses and dynamically by chaos.  We need
some framework to deal with the dynamical change of relationships
among elements.

\section{Logic for dynamic many-to-many correspondences}

Let a system be composed of many dynamic elements,
and chaotic motion be assumed in the system.  Then any
perturbation put in one element can be transmitted to other elements with
amplification.  In this situation, a chain of causal relationships
can bring about unexpected results \cite{Tsuda-book,Kawanabe}.
(A Japanese proverb for  such a `strong' causal connection is
``If the wind blows strongly then (finally) bath tubs sell well"
\cite{Tsuda-book,Kawanabe}.) The reasoning is the following:
1. If the wind blows strongly, the number of blind  people increases
due to the dust entering in their eyes.
2. If so, they try to earn  money by playing  a `shamisen' ( a
traditional Japanese musical instrument made of cat's skin).
3. If so, the demand for `shamisen' increases. 4. If so, cats are
hunted recklessly. 5. If so, cats extremely decrease in population.
6. If so, mice increase in population. 7. If so, (wooden) tubs are
gnawed by mice. 8. If so, tubs are sold well.

One of the authors(KK) encountered such situations when working with
coupled map lattices \cite{CML} or networks of chaotic elements \cite{KK-GCM}.
In a coupled map lattice with chaotic dynamics,
a tiny perturbation at a lattice point is amplified
to nearby elements.
A macroscopic order corresponding to the dissipative
structure\cite{Prigogine} can appear, but again be destroyed by the chaotic
dynamics until a next ordered structure appears.
In a network of chaotic elements,
clusters of synchronized oscillations may appear.
Identical elements can differentiate
due to the orbital instability in chaotic dynamics.
This mechanism, called (dynamical) clustering, is commonly seen
in globally coupled dynamical systems. Futhermore, the members of a
strongly correlated group change in time,
leading to a ceaseless change of relationships (see  {\bf Kaneko} in the
present volume).  Clustering and collective behaviors can
also be seen in globally coupled oscillators as
studied by {\bf Nakagawa}.  The synchronization between external (limb's)
and internal (neural) oscillations is essential to the model of
bipedal locomotion by {\bf Taga} in this volume.

Besides in chaos research \cite{Aizawa}, the use of one-to-one
correspondence has been challenged in many branches of science.
In brain science, the hypothesis of a grandmother cell has been
doubted.  Such doubts have led to the notion of a distributed representation
of information, and the research of
neural networks related with the spin glass theory.  However, such challenges
remain at a static level.  In contrast with the static logic, the necessity of
a dynamic logic has been postulated by
Malsburg, Vaadia, Aertsen, Dinse, Freeman, one of the authors (IT), and so on
( see the papers by {\bf Tsuda}, {\bf Aertsen}, {\bf Dinse}, and
{\bf Freeman} in the present volume).

In an ecological system the necessity of a logic {\sl that grasps a
complex system without reduction to an ensemble of simple elements}
has been stressed by Elton \cite{Elton}. Kawanabe has pointed out the
necessity of a logic to represent the above Japanese proverb.
Indeed it is known that in some ecological systems there
are  keystone species, a removal of which strongly damages the whole
ecosystem \footnotemark.

\footnotetext{This does not mean that such ecosystems are dominated by
(few) keystone species.  A role of a keystone species
implicitly emerges within an ecological network, through the amplification of
tiny causes as illustrated in the Japanese proverb.}

Ikegami and one of the authors(KK) have studied a population dynamics model
with many types of hosts and parasites, which are subjected to mutations.
In a weak coupling
regime, a one-to-one relationship between a host and a parasite holds, while
dynamic many-to-many relationships between pairs of hosts and parasites
emerge in a strong coupling regime, together with the maintenance of
a high mutation rate.
%The ecology resulting from the model gives a specific example for dynamic
%many-to-many correspondence.
We note that the resulting ecology is dynamically stable, sustained
by a high-dimensional chaotic state, in contrast with the
strong instability in a low-dimensional chaotic population dynamics.
Some theory for dynamic many-to-many correspondences
is required to allow for the diversity in an ecosystem.

In the present proceedings, {\bf Yomo} presents the dynamic clustering
of E-coli, bacteria.  Even if these bacteria have identical DNA, they
dynamically differentiate.  The one-to-one correspondence between
a genotype and a phenotype is invalid here.
A novel mechanism for the differentiation
of cells is proposed, based on the idea of dynamical
clustering (see {\bf Kaneko and Yomo}).
%The cell society is viewed as a system that follows
%a logic of dynamic many-to-many correspondence.

Let us recall the history of Japanese literature.
About three hundred years ago, it was popular to have
ceremonies during which  Haiku's (short poems) were recited.
The ceremony staged poets who made poems in succession,
following the previous poem by somebody else.  A poet ``interprets" the
previous poem by him(her)self. This interpretation, of course,
may be different from the original poet's.  Thus
mis-interpretation is enhanced successively, but as a whole
the sequence of poems forms some art more aesthetic than that created by
a single poet.  This process consists of the dynamic amplification of
small deviations.  ``Collective" art at a higher level emerges
as an ensemble of poems.  One might think that this process is just
a kind of bottom-up approach to collective art.
This is not necessarily true.  To address this problem, let us
re-examine the top-down and bottom-up approaches.

\section{Top-down, bottom-up, and emergence}

There have been long debates between the top-down and bottom-up
proponents in artificial intelligence and neural networks.
In both approaches, it is assumed implicitly that the top level is
represented by a few degrees of freedom, while
the bottom level may involve a huge number of  degrees  of freedom.
In the bottom-up approach some kind of ``order parameter"
constructed from the lower level is viewed as a representation at a higher
level, related with some macroscopic behavior. In the top-down approach
only a few instructions are sent as messages to lower-level elements.

As a natural compromise between these approaches, the
inclusion of a weak feedback between the top/bottom levels
has been proposed.
An example is given by a simulation of ants with pheromone \cite{ant}.
In the simulation, an ant emits pheromone when it has food, while
other ants are attracted by pheromone through their motion.
A collective field of pheromone is  formed by the ants' motion.
Since the dynamics of a lower-level unit (an ant's motion)
is governed by the higher-level dynamics ( collective field of pheromone),
this scheme is
analogous with the Prigogine's dissipative structure \cite{Prigogine}
or Haken's slaving principle \cite{Haken}.

In these approaches the relationships between elements
are fixed.  Although it may be possible to introduce
nontrivial dynamics (e.g., in the ants' motion or in the field of
pheromone), the behaviors of each element are passive and
totally susceptible to a higher-level.

Most papers in the present proceedings adopt
a different approach in the following senses. First, the top level is not
necessarily
represented by a few degrees of freedom; second, the relationships between
elements at a lower level can often change dynamically. At  first glance,
the first point may just look like a complication.  This
is not necessarily so.  Even in the midst of highly disorganized states,
ordered motion governed by a few degrees of freedom often
emerges, which, however, does not last for ever due to the second point
(the dynamic change of relationships). Again, high-dimensional motion
comes back, until another structure emerges.  This mechanism, called
{\sl chaotic itinerancy} \cite{KK-GCM,Ikeda,Tsuda-CI,Davis}, can replace the
views
of the top-down and the bottom-up
approaches.  In a network of chaotic elements, for example,
the order at the top level is destroyed by a chaotic revolt against the
slaving principle \cite{KK-GCM}, in contrast with passive elements
in traditional approaches. In the dynamic neural network model by one of the
authors (IT), the chaotic itinerancy leads to a spontaneous recall of memories
\cite{Tsuda-CI}.  A similar dynamical behavior is
also observed in real brain activities.
In the present proceedings, this topic is studied in the papers by
{\bf Tsuda}, {\bf Aertsen}, {\bf Dinse}, {\bf Freeman}, and {\bf Nozawa}.

In the population dynamics model mentioned in
\S 2 \cite{homeo}, the higher-level corresponds to the collective dynamics
for survival as an ensemble of many types.  This higher level emerges from the
bottom level, but
it is not necessarily represented by a few degrees of freedom.
An ecosystem like {\bf Ray}'s TIERRA, with species of programs, is
not necessarily represented by a few sets of features.
Species with different properties appear successively
in a specific condition.  In the paper by {\bf Palmer},
a stock market is formed as a higher level.

The term ``emergence" is often used as spontaneous appearance
of an upper level description
without explicit instruction for it \cite{Polanyi}.
If the upper level is  represented by a few order parameters,
the term ``emergence", in this
case, is just a rephrase of a dissipative structure
in nonequilibrium statistical mechanics, or a  collective behavior in
equilibrium phase transitions.
When the upper level is not represented
by a few degrees of freedom, however, the emergence is no
longer trivial.

The term emergence is often used when the behavior found in a computer
experiment is not written in a model explicitly as an algorithm.
In order to be an ``emergent" behavior, an explanation of it
from the implemented program should require at least as
much as the information as the direct computation.
Such ``wishes" for the construction of emergence
may not be rigorously accomplished as long as one uses a finite-state machine
(e.g., digital computer) for a finite time interval,
since the simulated behavior is obtained by a completely controlled program,
with a finite amount of information.  As long as all the information
is finite, it is difficult to define a behavior ``unexpected"
from the implementation.
Thus many people have tried to avoid this term in the workshop,
although we  tacitly feel that
``emergence" is necessary for the understanding of
the dynamics in brains and in biological
evolution.

There can be two possibilities to remedy the above impossibility of
emergence.
One is the assumption of
the use of infinite cells or tapes and/or a possible use of infinite
time step computations.  In connection with the undecidability of the halting
problem, it may be possible to have emergence by taking an
infinite-step limit.

The other is the introduction of uncontrollability
up to an infinite precision.  Let us recall the Japanese proverb;
given the strong wind, the outcome that the tub is sold well is
rather unexpected.  To explain this process, we have to follow each step of the
reasoning, which itself is easily affected by a small error.
The outcome, in this case, can be emergent behavior.
By the introduction of an analog computer with chaotic dynamics and/or
error in it, one may thus expect the occurrence of emergence. We note that
notions of computability in a real-number machine, discussed
by \cite{Blum,Moore} in connection with chaos, may be essential to
explore this possibility.

Another way for the introduction of uncontrollability may be
the introduction of a quantum mechanical computer,
as is discussed by {\bf Conrad}.

Even in our digital computer of finite resources, there can be some hope.
Chaos, for example, cannot be simulated rigorously by
any digital machine:  As long as a state in the machine is
finite, the dynamics becomes periodic (Poincare recurrence),
finally.  Still, we can  grasp the features of chaos
(by taking the limit of infinitely many states) in a digital
computer.  In a similar manner,
emergence may be defined by taking the limit of infinitely many states
from our digital machine.
To understand the nature of this limit, we need to make more efforts to
construct a model with some kind of emergence, and also
some mathematical studies on the relationships between digital and other
computers.

\section{ From a Descriptive to Constructive approach of Nature}

Structural stability \cite{Smale}
had been presented and recognized as a necessary
condition of a real model, before the significance of chaos was appreciated.
On the other hand, for a system with structural instability, a slight change of
the
model may lead to behaviors with different characteristics from real
solution's.
This is why such a system is believed not to be
a good model for nature. However, structural instability can
widely be seen in a nonlinear system including chaos.  For example,
chaos in the logistic map $x'=ax(1-x)$ cannot exist in an open interval
in the parameter space $a$. This means that a map with some fixed parameter
has no topological equivalence in any neighborhood of that map, thus implying
structural instability.

Chaos may have  another transcendental nature.
In some cases with non-uniform hyperbolicity, chaos may lack
the pseudo-orbit-tracing property\cite{POTP,POTP2}. If so, this
means that individual
orbits in chaos cannot be traced by experiments or by numerical
simulations.
It is still questionable if a whole attractor, i.e., a strange attractor
itself can be traced in experiments.
 At least one counter-example exists against the assertion that
a strange attractor itself is traced. In some chaotic systems
such as the Belousov-Zhabotinsky reaction map,
chaos looses characteristics such as topological and measure-theoretic
quantities,
affected by noise, and consequently order that
does not exist in any neighborhood of the original system appears
\cite{Matsumoto and Tsuda}.
A drastic change appears in the case of somewhat large noise, but the
calculation
of the Kolmogorov-Sinai entropy implies
the creation of different chaotic systems even in case
of infinitely small perturbations.
This so called noise-induced order \cite{Matsumoto and Tsuda}
has been interpreted in terms of an observational mismatch between the system's
inherent observation window, i.e. Markov partition, and the external
observation window forced by noise.

A complete description of chaos needs an infinite amount of information.
By a slight change of coding, in the case of a chaotic system,
the description of the system as, for instance, a finite automaton can change
drastically.

{\bf Crutchfield} has dealt with chaos as a class of various levels of
finite automata.
The input information is hierarchically classified as a language
accepted by a machine
that is constituted
of chaos with a finite observation window and observed symbol sequences.
Then, chaos appears as a kind of a finite automaton according to the respective
observation window.  One of the authors(KK) also discussed the
dynamics of  a coding scheme in a network of chaotic elements, where
the coding tree of observed symbol sequence changes forever \cite{KK-GCM}.
Any difference in the observation precision
leads to a crucial change of the dynamics \cite{KK-GCMINF}.

Thus chaos manifests itself in various forms, depending sensitively on
its description.
By these observations, one of the authors (IT)
has introduced the term ``descriptive instability",
although further studies are necessary for its mathematical
definition.

Noise-induced order, a coding tree in the network of chaotic elements,
and Crutchfield's $\epsilon$ -machine have introduced novel viewpoints with
regard
to the `observation' or `description' in complex systems.
Thereby, one may notice the need of a more extended concept than structural
(in)stability in order to capture legitimately all the features of
complex systems. Here, what we need is not a concept representing
the system, but a concept about an `observer' describing the system.
Noise-induced order means that a change of the description of chaos
brings about a distinct phase of chaos due to the ``descriptive instability"
of chaos.

The notion of ``descriptive instability" raises a question about
the validity of a descriptive way of modeling.
Since a one-to-one correspondence is not possible for
each elementary process, a descriptive approach that always accompanies
analysis is not always
relevant for the understanding of a complex system.

In physics, we are used to adopt a descriptive
approach;  for example, an equation at a macroscopic level
( like the Navier-Stokes equation) is approximately
derived from a microscopic level (like the Newtonian equation of
many particles), and then numerically simulated.
Conventionally, a model equation in physics is believed to have a one-to-one
correspondence with the phenomenon concerned.

Studies on chaos, however, may lead one to question this
traditional picture of nature.  Let us take the example of chaos in
fluid dynamics.
If one carries out a splendid numerical simulation on sets of
equations with the velocity and temperature fields ( e.g., Navier-Stokes
equation with buoyancy and heat), one possibly can get the same oscillatory
behavior
of rolls as in experiments. Does this success give any better intuition
on the origin of this strange oscillation than that
provided by a simple chaotic system?   The authors think
that the answer is ``No" for most scientists who know about chaos.
One of the most important lessons
from chaos lies in that it has opened the road to a qualitative dynamical
viewpoint.  Low-dimensional chaos can provide a universal mechanism
underlying the onset of turbulence.

By developing further the viewpoint of chaos, the importance of
a constructive, rather than a descriptive, approach  has been pointed out.
An example of such a constructive approach is the coupled map
lattice \cite{CML},
proposed by one of the authors (KK) for the studies of
spatiotemporal chaos, pattern dynamics, and so on.
The model, constructed by combining some basic procedures (
such as local chaos, diffusion, flow, $\cdots$),
cannot be derived from a
first-principle equation like in conventional physics, but it
still has a strong predictive power
for novel phenomenology classes in complex dynamical systems.

A model cannot be exactly the same as nature herself anyway.
By ``descriptive instability" there may not be a well-defined
quantitative ``distance" between a model and nature.  Thus a descriptive model
based on microscopic knowledge is not necessarily quantitatively very close
to the phenomena under consideration.  Even if a descriptive model
happens to be
quantitatively close to nature, in a complex system, it
is in principle intractable to check
detailed correspondence between the model and nature,
numerically or experimentally.
Furthemore we often do not need the detailed information of nature
which is sensitive to the details of the models.
Rather, we are more interested in universal aspects robust against
changes of the model.
In other words,
we go up to a higher-level description which focuses on structurally stable
aspects. Thus a  qualitatively correct model which forms some universality
class\footnotemark is strongly required, for which
the constructive approach is
often more powerful.

\footnotetext{Here we use the term ``universality class" as a
qualitative class,  thus, in a broader sense
than adopted in statsitical mechanics.}

Through the constructive approach, one tries to understand how such
phenomenology is legitimated, how large the universality class to be described
by phenomenology is,
and what the essence of the phenomena is. Only through this approach we can
see why some type of complex behavior is common in nature,
irrespective of
the details, and then we can predict what class of systems leads to such
behavior.

\section{Why Artificial Reality}

The constructive approach in the last section implies the
necessity of the construction of a model with artificial reality.
The behavior of a model is not
easily derived analytically in complex systems.
One needs computers as a
heuristic tool, as a hypothesis generator, rather than as a descriptor.
The activities often called ``artificial life" belong to this class
of modeling.

Such modeling is especially necessary when one deals with
historical phenomena, like evolution, since it is rather difficult to
understand one historical path, without knowing other could-be paths.
Construction of a model with artificial reality provides
an alternative approach when the traditional one faces difficulties.
In the present proceedings, papers by
{\bf Ray}, {\bf Hogeweg}, {\bf  Lindgren}, {\bf Suzuki},
{\bf Ikegami}, and {\bf H\"{u}bler} present successful examples, as well as
the report by {\sl Fontana} in the workshop \cite{Fontana}.
{\bf Palmer}, and {\sl Yasutomi} \cite{Yasutomi} have shown the power of the
artificial reality approach in economics.

Frequent criticism raised to the artificial modeling is the
lack of quantitative predictions.  In natural science, it
is often presumed to be ideal to predict quantitative results
obtained in quantitatively specified experimental conditions.
Such a precise quantitative prediction is not available in
artificial modeling. Still, the phenomena observed in the artificial
reality can provide a metaphor for what occurs (has occurred) in nature
and in human society.  Furthermore we can understand the
essence of ``real" phenomena through the artificial world, which
makes  qualitative prediction possible in a much broader sense.

In the present volume {\bf Li} discusses an expansion-modification system,
a kind of cellular automata with a growing number of cells.
The long-range correlation found in this ``artificial model"
made him and one of the authors (KK) to expect the existence of
long-range correlations
in real DNA sequences, which was later confirmed.
Another example demonstrating a possible connection between artificial and
``real" biology can be found in the paper by {\bf Kaneko and Yomo} in this
volume.

The significance of such ``artificial science" was first
pointed out by Simon, in the context of engineering science \cite{Simon}.
The  present constructive approach to complex systems also
has some applications to engineering problems, mainly
in the area of information processing.  In such a case, one
constructs a system by combining procedures.  Here we should note
that the combination often leads to some (emergent) performance
un-expected from the sum of procedures.  We have some examples in
these proceedings:
{\bf Nozawa} has shown that a combination of a
network of chaotic elements and
a neural network of Hopfield's type brings about
remarkable efficiency in optimization problems.
{\bf Kitano} has demonstrated that
the combination of a genetic algorithm and an L-system for developmental
processes
provides high efficiency in learning.
{\bf Taga} gives a beautiful application of the synchronization to
bipedal motion, while
{\bf Kopecz} shows some emergent performance in robots whose motion is
controlled by artificial
neural nets, so that they successfully avoid obstacles.

\section{Methodological problems in complex systems}

As we have discussed so far, the behavior in an
artificial world cannot be represented by
reduced sets of degrees of freedom while exactly corresponding to the
complex world.  We need novel methodologies to understand the complex
behaviors emerging in artificial models.  So far we do not have
established methods such as those in (equilibrium) statistical mechanics.
We discuss briefly some possibilities.

a) Multiple viewpoints:

It is often required to describe the observation from
many points of view.  For example, the understanding of
pattern dynamics in spatiotemporal
chaos requires both the views from real space and phase space.
Integration of dynamical systems theory and computation theory
is relevant to the understanding of  complexity in
chaos or cellular automata, as  shown by
{\bf Crutchfield} and {\bf Mitchell}.
For game dynamics, we need viewpoints both from an
algorithmic level of strategy, and dynamical systems theory,
as is seen in the studies of {\bf Lindgren}, {\bf Ikegami},
{\bf H\"{u}bler}, and {\bf Suzuki}.

b) Mathematical anatomy in a high-dimensional phase space:

In low-dimensional dynamical systems, ``anatomical" methods
of geometric structures in
phase space have been developed. In complex systems, it is required to
extend such anatomical studies to high-dimensional cases.

In a system with a static complexity such as the spin glass problem,
the anatomy of the (energy/fitness) landscape has clarified its ruggedness
\cite{SG}, while, for a dynamic case, the studies are more difficult,
(since ``anatomy" itself is a static tool),
although some pioneering approaches have recently been proposed
\cite{Shinjo,Ohmine,Konishi} in Hamiltonian dynamical systems with
many degrees of freedom ( e.g., in molecular dynamics).
The anatomical studies
exploring the high-dimensional phase space will be important in
systems with evolution and adaptability.

c) Naturalists' viewpoints

In complex systems, one possible direction is to make a
collection of complex behaviors, list them up  and then classify them.
This approach, borrowed from natural history, has been adopted
in complex systems.  The classification of the behaviors of cellular automata
by Wolfram \cite{Wolfram}, although not complete, can belong to this approach.
Physicists' preference to make a ``phase diagram" is the simplest
version of such classification.  In complex systems, we
have to face
more complicated classifications. The naive use of a phase diagram may not
be powerful, there.  Indeed, in cellular automata, the
lack of a continuous parameter
makes it difficult to construct a phase diagram following the
classification ( see also the paper by {\bf Mitchell} in the present volume).
If a system's dimension is very high,
construction of a phase diagram is not practical, where
one has to resort to more heuristic approaches.  In such cases
a naturalist's approach may be useful.

The three approaches mentioned above are not necessarily sufficient for
understanding all complex systems.
These are apparently the approaches from without\footnotemark,
which is usual and common to
conventional sciences. A constructivistic approach is different. A
constructivist tries to make an uncontrollable world inside a computer which is
controllable from outside,
to make the world be functional. A decisive point for the success of the
modeling lies in the construction of an internal mechanism.
Therefore, a constructivist's approach is inevitably an approach from within.
We discuss this point in the next section.

\footnotetext{Here we use the word ``without" as the antonym of ``within".}

\section{Internal observer}

In a system with artificial reality, one has
to construct an internal observer; otherwise
a system can never be intelligent. In molecular biological systems,
among others, Conrad\cite{Conrad}, R\"{o}ssler \cite{Roessler},
Matsuno\cite{Matsuno}, and one of the authors (IT) \cite{Tsuda-Endo}
have pointed out the significance of an
internal observer which reacts with high efficacy. In brain modeling
also, it has been pointed out that the introduction of an internal viewpoint
would be essential to understand `the brain understanding itself'. If chaos
works in many phases of brain activities, chaos can be a candidate of such an
internal observer. This viewpoint has been called ``chaotic hermeneutics" by
one of the authors (IT) \cite{Tsuda-Endo}.

The significance of internal viewpoints for the ``understanding" of systems
was first proposed by G\"{o}del \cite{Goedel} in
constructing a theory to involve a description, from without, of formal system
into again the formal system, thus a description from within. By this
constructive approach, G\"{o}del succeeded
to prove that there exists a theorem which
is true, but unprovable only by using theorems of the system. Hereby,
the complexity of formal systems was elucidated.

{\bf R\"{o}ssler} has proposed a new scientific paradigm, that is,
endophysics,
generalizing the internal viewpoint in a formal system to that in physics,
chemistry and biology.
In endo-world, it could be that the observation {\sl from within} differs from,
even contradicts to the observation {\sl from without}. Only the latter
observation has been
explicitly performed in conventional science. In complex systems, however,
as has been discussed here, the observation and description from without
are apparently insufficient. Hence, one may well have to take the
endo-viewpoint
for a sufficient understanding of complex systems.
{\bf R\"{o}ssler} has constructed chaotic models to recognize the world from
within in
these proceedings.

Through the approaches so far we are trying to make a
reconstruction of the story for complex systems.  This
reconstruction is not necessarily unique.
Since the stories can include many variables and parameters,
it is not possible to conclude that only one of them provides the
best model. Thus it could be said that
Ockham's razor has lost its edge for complex systems we face now.
{\bf R\"{o}ssler}'s talk in the workshop seemed to absolve
Ockham's razor.

{\sl acknowledgements}

The speculation here is strongly stimulated by the discussions during
Oji Seminars.  We are grateful to all the active participants in the
workshop, and also to those who contributed e-mail discussions for
the 1992 complex systems workshop at Kyoto.
We would also like to Frederick Willeboordse for critical reading of the
manuscript.  Last, but not least of course,
we would like to thank again the
Fujihara Foundation and the Japan Society for the Promotion of Science
for their generous supports for the workshop.


\begin{thebibliography}{999}
\bibitem{00}
All the papers in the present volume are cited with the
speaker's name in the workshop, which agree with the
first author of the names in the present volume,
except three cases:  Yomo's presentation is given by
the paper of E.Ko, T.Yomo, and I. Urabe.  An additional paper
by Kaneko and Yomo is inserted, since it is related with
the experimental paper by Ko et al.  Michael Conrad's presentation
is given by the paper of J-C. Chen and M. Conrad, while
H\"{u}bler's presentation is given by the paper by
A. Pierre, A. H\"{u}bler, and D. Pines.
Dinse, who was invited but could not attend, is so kind to
submit his could-be presentation in the volume.
\end{thebibliography}
\end{document}